\documentstyle[psfig]{mn}

\newif\ifAMStwofonts

\ifoldfss
  \ifCUPmtlplainloaded \else
    \NewTextAlphabet{textbfit} {cmbxti10} {}
    \NewTextAlphabet{textbfss} {cmssbx10} {}
    \NewMathAlphabet{mathbfit} {cmbxti10} {} 
    \NewMathAlphabet{mathbfss} {cmssbx10} {} 
  \fi
  \ifAMStwofonts
    \ifCUPmtlplainloaded \else
      \NewSymbolFont{upmath} {eurm10}
      \NewSymbolFont{AMSa} {msam10}
      \NewMathSymbol{\upi}     {0}{upmath}{19}
      \NewMathSymbol{\umu}     {0}{upmath}{16}
      \NewMathSymbol{\upartial}{0}{upmath}{40}
      \NewMathSymbol{\leqslant}{3}{AMSa}{36}
      \NewMathSymbol{\geqslant}{3}{AMSa}{3E}

    \fi
  \fi
\fi 

\ifnfssone
  \newmathalphabet{\mathit}
  \addtoversion{normal}{\mathit}{cmr}{m}{it}
  \addtoversion{bold}{\mathit}{cmr}{bx}{it}
  \newmathalphabet{\mathbfit} 
  \addtoversion{normal}{\mathbfit}{cmr}{bx}{it}
  \addtoversion{bold}{\mathbfit}{cmr}{bx}{it}
  \newmathalphabet{\mathbfss} 
  \addtoversion{normal}{\mathbfss}{cmss}{bx}{n}
  \addtoversion{bold}{\mathbfss}{cmss}{bx}{n}
  \ifAMStwofonts
    \ifCUPmtlplainloaded \else
      \UseAMStwoboldmath
      \makeatletter
      \new@mathgroup\upmath@group
      \define@mathgroup\mv@normal\upmath@group{eur}{m}{n}
      \define@mathgroup\mv@bold\upmath@group{eur}{b}{n}
      \edef\UPM{\hexnumber\upmath@group}
      \new@mathgroup\amsa@group
      \define@mathgroup\mv@normal\amsa@group{msa}{m}{n}
      \define@mathgroup\mv@bold\amsa@group{msa}{m}{n}
      \edef\AMSa{\hexnumber\amsa@group}
      \makeatother
      \mathchardef\upi="0\UPM19
      \mathchardef\umu="0\UPM16
      \mathchardef\upartial="0\UPM40
      \mathchardef\leqslant="3\AMSa36
      \mathchardef\geqslant="3\AMSa3E
    \fi
  \fi
\fi 

\ifnfsstwo
  \DeclareMathAlphabet{\mathbfit}{OT1}{cmr}{bx}{it}
  \SetMathAlphabet\mathbfit{bold}{OT1}{cmr}{bx}{it}
  \DeclareMathAlphabet{\mathbfss}{OT1}{cmss}{bx}{n}
  \SetMathAlphabet\mathbfss{bold}{OT1}{cmss}{bx}{n}
  \ifAMStwofonts
    \ifCUPmtlplainloaded \else
      \DeclareSymbolFont{UPM}{U}{eur}{m}{n}
      \SetSymbolFont{UPM}{bold}{U}{eur}{b}{n}
      \DeclareSymbolFont{AMSa}{U}{msa}{m}{n}
      \DeclareMathSymbol{\upi}{0}{UPM}{"19}
      \DeclareMathSymbol{\umu}{0}{UPM}{"16}
      \DeclareMathSymbol{\upartial}{0}{UPM}{"40}
      \DeclareMathSymbol{\leqslant}{3}{AMSa}{"36}
      \DeclareMathSymbol{\geqslant}{3}{AMSa}{"3E}
    \fi
  \fi
\fi 

\ifCUPmtlplainloaded \else
  \ifAMStwofonts \else 
    \def\upi{\pi}
    \def\umu{\mu}
    \def\upartial{\partial}
  \fi
\fi

\title{A nuclear grand-design spiral within the normal disc spiral of 
NGC~5248}
\author[S. Laine et al.]
{S. Laine,$^{1}$\thanks{Visiting Astronomer, Canada--France--Hawaii Telescope 
(CFHT).}\thanks{Guest investigator of the RGO Astronomy Data Centre}
J.H. Knapen,$^1$$^{\star}$ D. P\'erez-Ram\'\i rez,$^1$
R. Doyon$^2$$^{\star}$ and D. Nadeau$^2$$^{\star}$\\\
        $^1$Department of Physical Sciences, University of Hertfordshire, 
        College Lane, Hatfield, Herts AL10 9AB\\
        $^2$Observatoire du Mont M\'egantic and D\'epartement de Physique,
        Universit\'e de Montr\'eal,
        C.P. 6128, Succursale Centre Ville,\\ 
        Montr\'eal (Qu\'ebec), H3C 3J7 Canada}
      
\date{Accepted 1998 October 28.
      Received 1998 July 15}

\pagerange{\pageref{firstpage}--\pageref{lastpage}}
\pubyear{1998}

\begin{document}

\maketitle

\label{firstpage}

\begin{abstract}

We report the discovery of trailing grand-design nuclear spiral structure
in the spiral galaxy NGC~5248.
Two tightly-wound red spiral arms, with widths less than 1 arcsec
($\sim$~75~pc), 
emanate at 1 arcsec outside the quiescent nucleus of the galaxy, and can be 
followed for
3 arcsec, appearing to end before the radius of the circumnuclear starburst 
``ring'', which is at 5--6 arcsec or around 400 parsec. Combining our near-infrared
Canada--France--Hawaii Telescope adaptive
optics results with traditional near-infrared and optical imaging, we
show that spiral structure is present in this galaxy at spatial scales
covering two orders of magnitude, from a hundred parsecs to 15 kpc. 
Comparison with a {\it Hubble Space Telescope} ultraviolet
image shows how the location of the circumnuclear star formation
relates to the nuclear spiral structure. We briefly discuss possible 
mechanisms which could be responsible for the observed nuclear
grand-design spiral structure.

\end{abstract}


\section{Introduction}

Little is known about the structures that govern the dynamics of disk
galaxies within the central few hundred parsecs. Colour index images, made by
combining two broad-band optical or near-infrared (NIR) images, allow us to 
study the dust lane morphology in the nuclear region, which is
important for constraining numerical models of gas dynamics in the 
circumnuclear region (see, e.g., Knapen et al. 1995a). Structures expected
in the nuclear region include trailing or leading spiral structure or 
bars (e.g., 
in M100; Knapen et al. 1995b), and star forming (SF) regions (e.g., Elmegreen
et al. 1997; Ryder \& Knapen 1998).

Nuclear spiral structure has been observed recently in a
number of galaxies. These cases include 
the discovery of small-scale flocculent dust spirals in the cores of NGC
278 (Phillips et al. 1996), NGC 2207 (Elmegreen et al. 1998) and others
(Carollo, Stiavelli \& Mack 1998). More chaotic dust lane structures
have been seen in {\it Hubble Space Telescope (HST)} WFPC2 images of
M51 (Grillmair et al. 1997), M81 (Devereux, Ford \& Jacoby 1997) and M87 (Ford et
al. 1994; Dopita et al. 1997). High resolution {\it HST} NICMOS and 
adaptive optics (AO) Canada--France--Hawaii Telescope (CFHT) NIR images 
have also revealed nuclear spirals (Regan \& Mulchaey 1997, 1998; Rouan et al. 
1998).

We report the first detection of nuclear
``grand-design'' spiral structure as seen in NIR AO images of 
NGC~5248, a galaxy classified as SXT4 by de Vaucouleurs et al. (1991; RC3). 
We compare our new data with UV, optical and NIR
images that cover a large range of size scales. We use a
distance of 15.4~Mpc to the galaxy ($v_{{\rm sys}}$ = 1153 km~s$^{-1}$; RC3), 
which implies that 1 arcsec corresponds to
75~pc. NGC~5248 has a circumnuclear starburst ``ring'' (Pogge 1989;
Maoz et al. 1996; Fig.~1), but its nucleus is quiescent, as evidenced 
spectroscopically (Storchi-Bergmann, Wilson \& Baldwin 1996; Ho, Filippenko 
\& Sargent 1997).
The circumnuclear
SF properties of NGC~5248 were recently studied by Elmegreen et al.
\shortcite{elme}, and the dynamics of the main spiral structure were
modelled by Patsis, Grosb{\o}l \& Hiotelis \shortcite{patsis}. 

\section{Observations}

The $J$, $H$ and $K$ band images were obtained in 1997 March
at the CFHT using
the combination of the AO system PUEO \cite{rigaut1}
and the Montr\'eal NIR camera MONICA (Nadeau et al. 1994). 
The total
integration time was 180 seconds in each band. Individual images were
sky-subtracted, flatfielded, corrected for bad pixels and combined into
a final mosaic. The uncorrected seeing was 0.6--1.0~arcsec, while the
corrected resolution is $\sim$~0.15 arcsec. We used throughout the
image scale of 34.38$\pm$0.07 mas~pixel$^{-1}$ and rotation prescription
from Rigaut et al. (1998b). The resulting orientation of the images is 
accurate to within $0\fdg 1$. The
total field of view of the images shown here is 12.4 arcsec~$\times$ 12.4
arcsec.  Colour index images were made by combining the images after
aligning the nuclear peak (in the absence of field stars). Although we 
were able to define the position of the nucleus accurately enough for
alignment, the $K$-band profiles across the 
nucleus in the original exposures are flat, causing an artificial blue 
nuclear core in the $J-K$ and $H-K$ images. For this reason we
do not discuss the area within the central arcsec of the $J-K$ image.   
The calibration of the images was performed in the standard way by
observing stars from the list of Landolt \shortcite{land}. The 
uncertainty in the calibration is about 0.05 mag. 

We also present NIR and optical images with a larger field of
view. The NIR images were taken with MONICA at the CFHT as part of a 
survey of galaxies with
circumnuclear regions of star formation (P\'erez-Ram\'\i rez \& Knapen 1998;
Knapen et al. 1999) and cover the central 1 arcmin $\times$ 1
arcmin with a resolution of 0.6 arcsec.
An optical $B$-band image from the data archive of the Isaac Newton
Telescope is also presented. The field of view is 10
arcmin $\times$ 10 arcmin, and the resolution 1.6 arcsec. Finally, we use
a Faint Object Camera image from the {\it HST}
archive, taken at an effective wavelength of $\sim$~2300~\AA~. The image 
covers the central 22 arcsec $\times$ 22 arcsec (Maoz et al. 1996).

\section{RESULTS}

\begin{figure}
\label{fig1}
\psfig{figure=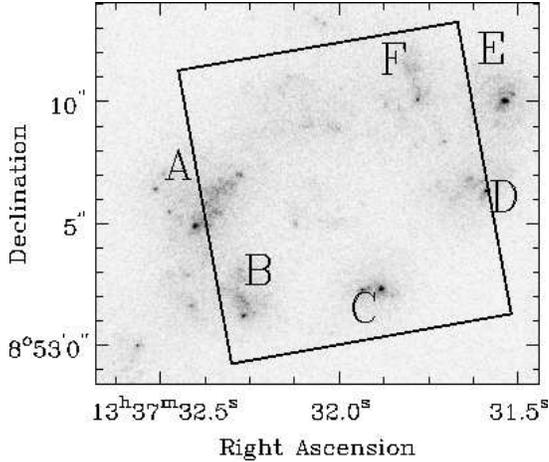,height=2.4in,angle=0}
\caption{An {\it HST} Faint Object Camera F220W  ultraviolet image of the
starburst ``ring'' in NGC~5248. The
coordinates are in J2000.0. Some of the SF sites are marked with letters
A--F for comparison with the NIR images presented in Fig.~2. The area 
that is covered by the CFHT images presented in Figs. 2 (left) and 4
has been marked with a box.}
\end{figure}

\begin{figure*}
\label{fig2}
\psfig{figure=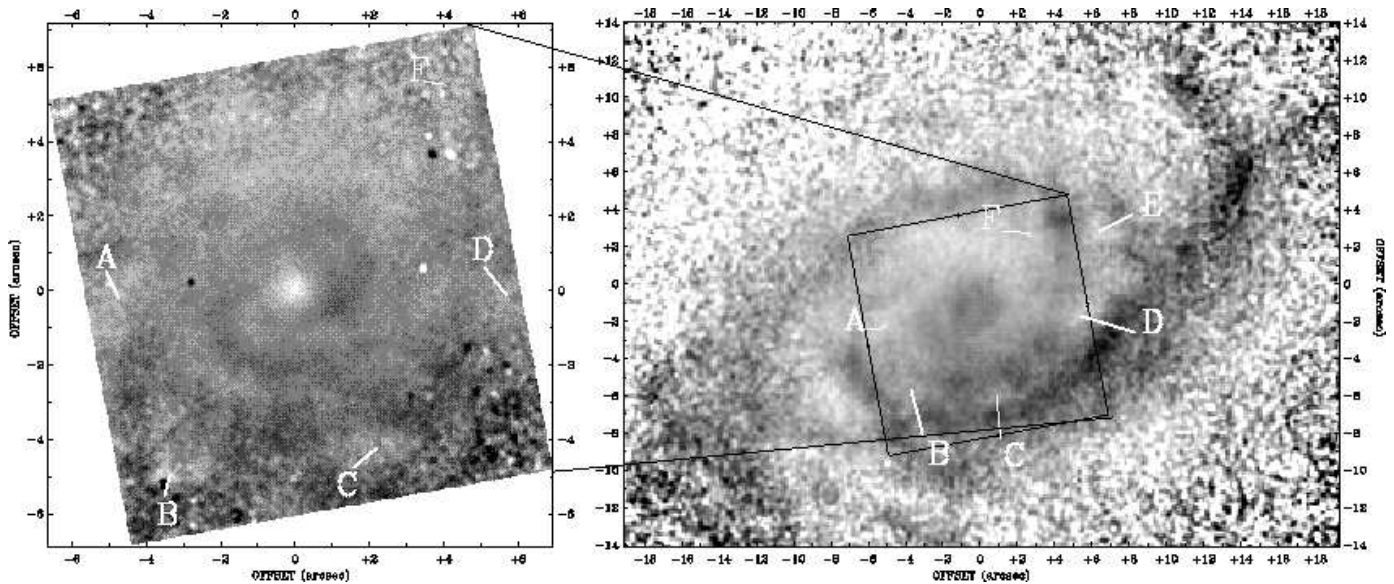,height=3.0in}
\caption{Left: CFHT adaptive optics $J-K$
colour index image of the core region of NGC~5248, showing the nuclear
spiral in darker shades. The range of colours is approximately from 0.6
(lightest regions) to 1.2 (darkest). Some of the SF sites have been marked
with capital letters to assist the comparison to the UV image (Fig.~1).
Right: $J-K$
colour index image of a larger area around the central region of
NGC~5248, as imaged in 1995 with the CFHT. The area of the nuclear image
is shown with a box. The lightest colours have values around 0.8 whereas
the darkest colours along the arms have values around 1.1. Again, some of
the SF sites have been marked with letters A--F. North is up and east to
the left in the images.}
\end{figure*}

The major axis diameter of the circumnuclear ``ring'' (CNR) of NGC~5248 
(Fig.~1) is $\sim$~11 arcsec, or 800 pc. The SF
``ring'' is incomplete and patchy in UV emission. Some of
the SF sites have been marked with letters A--F in Figures 1 
and 2.
Some weak and dilute emission can be seen
within the nuclear area at radii within the ``ring'', but no SF
spiral structure is visible, as confirmed by Isaac Newton Telescope archival 
H$\alpha$ images (not shown here).

Figure 2 (left) shows our AO $J-K$ colour index image. Some of the
SF regions of the circumnuclear ``ring'' are partly visible near the edges 
of the image, and
show up as whitish patches. The most interesting feature in
the image is the red (dark in the figure) spiral structure. Two arms
with widths less than 1 arcsec (75 pc) originate at one arcsec northeast 
and southwest of the nucleus, and can be followed for 3 arcsec. They
appear to end
before the radius of the CNR (which is 5.5 arcsec). 
The two-armed spiral structure is trailing (assuming the
spiral arms in the outer disk are trailing). Its red colour suggests that 
it is a dust spiral. We emphasize that the colour variations across the 
image are not large (Table~\ref{tbl-1}). The contrast between the nuclear
spiral and the regions immediately surrounding it is around 0.05 mag.
The same two-armed spiral can be seen in $J-H$ and $H-K$ images, but
it is most obvious in the $J-K$ image.

\begin{table}
 \caption{Colour indices of the structures within the circumnuclear ring
 in magnitudes.
 The relative uncertainties are smaller than 0.03 mag. \label{tbl-1}}
 \begin{tabular}{@{}lcc}
 \hline
 Structure & $J-H$ & $H-K$ \\
 \hline
 Nucleus & 0.72 & \\
 Nuclear spiral & 0.85 & 0.15 \\
 Star formation ``ring'' & 0.79 & 0.11 \\
 \hline
 \end{tabular} 
\end{table}

Figure 2 also shows the relation of the nuclear spiral structure to the
outer disk of the galaxy. The right panel of Fig.~2 is the CFHT MONICA
$J-K$ colour index image, made without AO. Only the inner part of the
image, covering an area of
38 arcsec $\times$ 28 arcsec, is shown here. Again, red (dusty) spiral
structure can be seen, most clearly just outside the nuclear starburst
``ring'', which is located very close to the edge of the AO image.  Finally,
in Fig.~3 we present the $B$-band image, showing the main optical spiral
structure in the disk of NGC~5248, at radii of several kpc.  Outside
the central 2--3 arcmin region, a set of fainter spiral arms can be
seen, extending out to at least 15 kpc (cf. Fig.~5 in Patsis et
al. 1997). 

Our broad-band AO $J$, $H$ and $K$ images show a variety of structure
caused by SF regions, most clearly seen in $J$ (Fig.~4)
where the contributions of emission from young stars, but also of dust
extinction, are largest. In all of the AO images, the strongest star formation
region is clearly seen in the ``ring'' to the east of the nucleus. The
position angle of the isophotes is 120$\degr$$\pm$10$\degr$, defined by
the starburst ring, but the central 4 arcsec (300 pc) region is rounder
than the region outside it. The FWHM of the nuclear peak is 0.7 arcsec.
The nucleus is well defined in all images, and no
double- or multiple-peaked structures exist.

Estimating the exact shape of the point spread function (PSF) in the
various broad-band images is a difficult issue in AO imaging, and since in
general different PSFs can give rise to artifacts in a colour index
image, we have used two other, independent, techniques to confirm the
existence of the red nuclear spiral. Both techniques were applied to the
AO $J$ band image. First, we used the unsharp-masking technique, whereby
the $J$ band image was smoothed to a resolution of 0.7 arcsec,
and then subtracted from the original, unsmoothed, image. This technique
may introduce artificial light and dark concentric circles, but our
resulting image (Fig.~5) clearly shows the nuclear
spiral arms as non-circular features. In a
second test, we fitted ellipses to the isophotes of the $J$ band
image, built a model from the fitted ellipses, and subtracted the model
from the original image. Again, the residuals reveal the nuclear spiral
arms. Our tests thus confirm that the red nuclear spiral arms, as
seen in the $J-K$ image, are not artifacts of the techniques used.

The colour index variations across the central 10 arcsec region are small
(Table~\ref{tbl-1}). Additionally, the residuals after subtracting a
model of fitted ellipses reveal that the depletions in surface
brightness in the $J$-band are only at about 5 per cent level when
compared to the surrounding regions. Therefore, the (red) nuclear spiral
is likely to be composed of dust and gas only, without a counterpart in
the old or young stellar distribution.

Although we do not observe the nuclear spiral arms to connect to the CNR 
from inside, red spiral
arms (right panel of Fig.~2) connect into the CNR from outside, and can be 
followed out to 10 arcsec radius. Near the locations
of the outer ends of these red spiral arms, the main optical spiral structure
begins,
delineated by the bright SF regions seen in Fig.~3. The main strong
SF spiral arms end near 70 arcsec from the nucleus, where Patsis et al. 
\shortcite{patsis} placed the dynamical 4/1 resonance. Beyond 
this distance, a set of fainter arms exists.
In deeper exposures the faint arms appear as extensions
of the main spiral structure (Chromey, Elmegreen \& Harrison 1995). The
gap, or the minimum in the spiral intensity, could well mark
the corotation radius (Patsis et al. 1997).

\section{DISCUSSION}

In general, the most common tracers of spiral structure are bright, young,
massive stars and the \hbox{H\,{\sc ii}} regions around them, most readily
detected in optical H$\alpha$ images (e.g., Sandage 1961). The underlying density
enhancements, best seen in NIR $K$-band images
where the effects of young stars and cold, extincting dust are much
smaller than in the optical regime, show a smoother and more continuous variation in intensity 
between the arms and
the interarm regions (e.g., Grosb{\o}l \& Patsis 1998). In this paper, 
we use different tracers of the
spiral structure. The patchy spiral structure seen in the optical image 
in Figure~3 is made up of emission from young stars and SF 
regions, which have bluer colours than the surrounding disk of older
stars. In contrast, the spiral structure seen in the NIR images is
redder than its surroundings, tracing the reddening presumably caused by cold
dust structures. In the $J-K$ image, the red colours may also have a small
contribution from hot dust which begins to show up in the $K$-band at
2.2~$\mu$m.

Red, grand-design nuclear spiral arms, as reported here, have not been 
observed at such small scales before. In contrast,
flocculent, dusty spiral arms and chaotic dust structures have been
seen in the cores of several disk galaxies, both in the optical and in 
the NIR (see references in Introduction). What causes the nuclear grand-design
spirals in NGC~5248? 

The main optical spiral structure of NGC~5248
(Fig.~3) was modelled by Patsis et al. \shortcite{patsis}.
According to them, the main spiral exists between the dynamical 
inner Lindblad resonance (ILR) and the inner 4/1 resonance. In their
model, the (outer) ILR occurs at a radius of 30--40 arcsec. In that case, the
starburst ``ring'' at {\it r}~=~5 arcsec is located very far inside the
radius of the outer ILR.
It is unlikely that the nuclear trailing spiral structure in NGC~5248 is
related to an inner ILR, as the spiral structure there 
would be expected to be leading (e.g., Knapen et al. 1995a; Combes 1996). 
High resolution
kinematical data, combined with dynamical modelling, is needed for a
definite study of dynamical resonance structures.

It is possible that a nuclear gas (and dust) disk exists around
the core of NGC~5248.
Small perturbations, such as massive gas clumps, will drive spiral 
structure in the gas disk, but it is unlikely that grand-design spirals 
can be generated by this
mechanism. No nuclear bar can be seen within the SF ``ring'', although the
spiral structure could still be driven by an outer oval distortion. Nuclear gas
disks have been seen recently in CO emission of some
galaxies (e.g., Sakamoto et al. 1995; Laine et al. 1999), and high resolution 
observations of gas
tracers in NGC~5248 are required to see if such a disk exists in
NGC~5248.

\begin{figure}
\label{fig3}
\psfig{figure=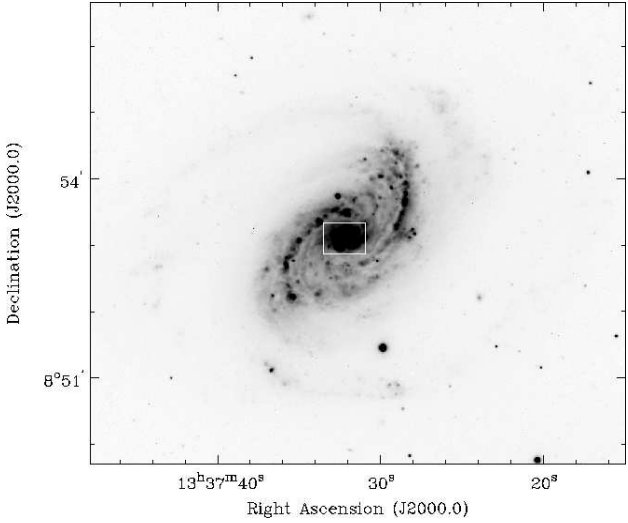,height=2.7in}
\caption{$B$ band
optical image of the disk of NGC~5248. A box indicates the area shown on
the right of Fig.~2, demonstrating how the different
scales interlock. Spiral structure is visible on all scales, from very
close to the nucleus to far out in the disk.}
\end{figure}

\begin{figure}
\label{fig4}
\psfig{figure=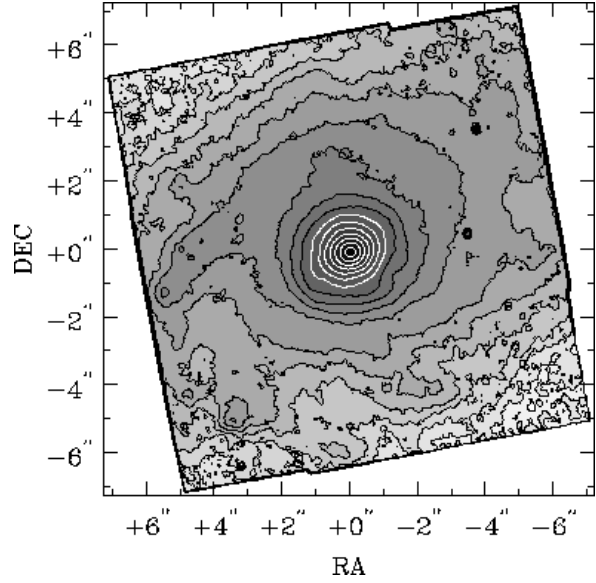,height=3.0in,angle=0}
\caption{Adaptive optics $J$-band image with about 0.2 arcsec
resolution as obtained at  the CFHT. The grayscale and
contour levels are from 13.5 to 16.9 mag~arcsec$^{-2}$, in steps of 0.2
mag~arcsec$^{-1}$. The contours near the nucleus have been drawn in
white to indicate the lack of spiral or other features in a broad-band
NIR image. The circumnuclear star formation ring shows up near the perimeter
of the image.}
\end{figure}

\begin{figure}
\label{fig5}
\psfig{figure=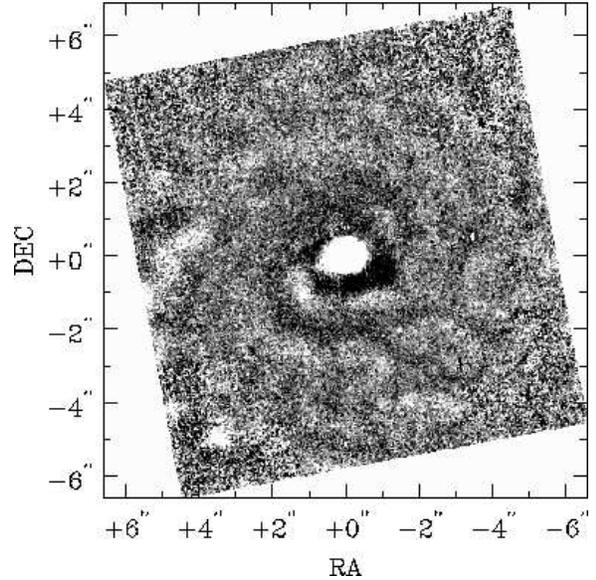,height=3.0in,angle=0}
\caption{Unsharp-masked image in $J$ band of the nuclear
12~arcsec x 12~arcsec region of NGC~5248.}
\end{figure}

Elmegreen et al. (1998) have suggested another possible mechanism
for creating nuclear spiral structure. Acoustic spirals, generated by
the amplification of sound waves at small galactocentric radii, may
produce spiral-like waves. However, acoustic waves are likely to produce
multiarmed spirals. Modes higher than {\it m}~= 2 are weak or absent in
our NIR images of the nuclear region of NGC~5248. Deeper imaging
of the centre of NGC~5248 must be obtained to see whether {\it m}~$>$ 2 spiral
structure exists, and thus whether acoustic 
spirals are a viable generating mechanism of the observed spiral
structure. 

Finally, it is possible that the spiral structure is continuous
from about 1 arcsec radius to the outermost optical spiral
arms. Recent modelling by Englmaier \& Shlosman (in preparation)
shows that a single ILR is capable of driving a two-armed spiral 
from corotation down to a few hundred pc or less from the nucleus. 
The CNR forms inside the single ILR because of the shock focusing
and subsequent action of self-gravity in the gas.
Deeper images with better signal to 
noise ratios are required to investigate the continuity of the
spiral structure in NGC~5248.

\section{Conclusions}

We have detected a pair of red spiral arms in the
central 100--300 pc region of NGC~5248.
The nuclear grand-design spiral structure in NGC~5248 is real, and
not an artifact of the observational techniques used. We have
used three different techniques (a colour index image, unsharp masking,
and subtraction of a model-based ellipse fit to the surface
brightness distribution) to isolate the nuclear spiral from our broad-band 
NIR AO images, and confirm the spiral with all three techniques.
It appears unlikely that the nuclear spiral is related to acoustic spirals. 
However, the dusty spirals may be an
indication that a weak non-axisymmetric component exists and drives the 
grand-design spirals in a mildly gravitating nuclear gas disk. Sensitive, high 
resolution observations of the molecular and
neutral gas would help to resolve the question of the origin of the
nuclear spiral structure.

\section*{Acknowledgments}

We thank the referee, Dr. John Beckman, for helpful comments. We are 
grateful to Drs. B. G. Elmegreen, J. Collett, 
P. Patsis, R. F. Peletier and I. Shlosman for their comments. We are
grateful to Drs. P. Englmaier and I. Shlosman for releasing results
before publication.
CFHT is operated by the National Research Council of Canada, the
Centre National de la Recherche Scientifique de France and the University of 
Hawaii.
Part of the data presented here is based on observations made with the Isaac 
Newton Telescope operated on the island of La Palma by the Isaac Newton Group
in the Spanish Observatorio del Roque de los Muchachos of the Instituto
de Astrof\'\i sica de Canarias. Part of the data are based on observations
made with
the NASA/ESA {\it Hubble Space Telescope}, obtained from the data archive at
the Space Telescope Science Institute. STScI is operated by the
Association of Universities for Research in Astronomy, Inc. under NASA
contract NAS 5-26555.

\label{lastpage}

\end{document}